\documentclass[%\
 reprint,
 amsmath,amssymb,
 aps,
]{revtex4-2}

\usepackage{lipsum}
\usepackage{color}
\usepackage{changes}
\usepackage{graphicx}% Include figure files
\usepackage{dcolumn}% Align table columns on decimal point
\usepackage{bm}% bold math
\usepackage{hyperref}% add hypertext capabilities
\hypersetup{
    colorlinks=true,
    linkcolor=blue,
    filecolor=blue,      
    urlcolor=blue,
    citecolor=blue
    }

\usepackage[capitalize]{cleveref}

\begin{document}

\preprint{APS/123-QED}

\newcommand{\aw}[1]{\textcolor{red}{[AW:#1]}}
\newcommand{\warn}[1]{\textcolor{blue}{#1}}
\newcommand{\error}[1]{\textcolor{red}{#1}}

\title{Quench dynamics of stripes and phase separation in the two-dimensional $t$-$J$ model}

\author{Luke Staszewski}
\email{lstaszewski@pks.mpg.de}
\affiliation{Max Planck Institute for the Physics of Complex Systems, N\"othnitzer Strasse 38, Dresden 01187, Germany}

\author{Alexander Wietek}%
\affiliation{Max Planck Institute for the Physics of Complex Systems, N\"othnitzer Strasse 38, Dresden 01187, Germany}

\date{\today}

%----------------------------------------------------------------------%
%  abstract
\begin{abstract}
We investigate the fundamental dynamical process of an initial quench of the chemical potential of the two-dimensional $t$-$J$ model. Depending on the ground state phase, sharply different dynamical behavior of the charge distribution and entanglement properties are observed. In the stripe phase, the intertwining of the spin and charge density waves remains stable under time evolution. A ballistic spreading of the charge density is observed with a propagation speed that is only weakly dependent on the coupling ratio, $J/t$. Moreover, in the phase-separated regime for large $J/t$, we report a complete dynamical freezing of charge degrees of freedom within, where even under long time evolution the entanglement entropy remains bounded. Our results are obtained by combining large-scale exact diagonalizations and matrix product state techniques for time evolution. 
\end{abstract}

\maketitle

%----------------------------------------------------------------------%
%  section: introduction
\section{Introduction}
\label{intro}

Non-equilibrium dynamics can serve as a window to probing the intricate physics of strongly correlated quantum matter.
While conventional transport measurements are often readily accessible in experiments, novel pump-probe experiments have
more recently yielded many novel insights into the physics of strongly correlated superconductors~\cite{Measson2014,Liu2020,Zhang2024}. Moreover, ultra-cold atom experiments allow us to closely track the dynamics of individual atoms~\cite{Schneider2012,Langen2015,Scherg2018,Brown2019}, and interesting dynamical processes, like quantum quenches~\cite{Langen2015,Mitra2018} and Floquet dynamics~\cite{Bukov2015}, can be studied in unprecedented detail.

A particularly fascinating class of quantum materials are doped Mott insulators. While the high-temperature copper oxide superconductors are prime examples where doping a parent Mott insulator can induce superconductivity, we now know many more compounds exhibiting similar physics. This ranges from the transition metal dichalcogenides~\cite{Shi2015,Manzeli2017} to twisted moir\'e materials~\cite{Cao2018,Yankowitz2019,Tang2020,Wang2020,Li2021,Ghiotto2021} and the novel nickelate superconductors~\cite{Li2019,Wang2024}. From a theory perspective, the Hubbard model and the related $t$-$J$ model~\cite{Anderson1987,Zhang1988,Emery1988} in two dimensions have been understood to capture the essential aspects of these complex phenomena - not without having posed significant challenges for both analytical and numerical studies~\cite{Qin2022,Arovas2022}. Modern developments in numerical algorithms have led to rather firm insights concerning the ground state~\cite{LeBlanc2015,Zheng2017,Qin2020,Huang2017,Huang2018,Wietek2022,Baldelli2023} as well as finite-temperature properties~\cite{Wietek2021,Wietek2021b,Schaefer2021}. Currently, non-equilibrium properties of these models are at the forefront of computational and theoretical research.

Early on, the dynamics of strongly correlated fermionic and bosonic systems was studied intensely in one spatial dimension~\cite{Manmana2007,Kollath2007,Cramer2008}, where analytical techniques, such as the Bethe ansatz as well as numerical techniques are both readily available. These studies revealed that nonthermal steady states which strongly depend on the initial condition can be realized after a quench in the system parameters. 

In two spatial dimensions, several studies have considered the dynamics of a single hole in a Mott insulating background~\cite{Kadow2022,Ichinose1992,Bohrdt2020}. Ultracold atoms have been used to actively study the dynamics of mobile holes in an antiferromagnetic background experimentally~\cite{Ji2021,Koepsell2019,Kale2022}, and mobility constraints arising from dipole conservation laws and fracton-like physics have drawn a theoretical interest to the subject~\cite{Pretko2020,Sous2020a,Sous2020b}. In particular, $t$-$J_z$ models on a square lattice with Ising interactions have been shown to exhibit exact dipole conservation leading to dynamical localization of a single doped hole while bound pairs of two holes are mobile~\cite{Sous2020a,Sanyal2024}.

%----------------------------------------------------------------------%
%  figure 1: cartoon of phase separation
\begin{figure}[b]

  \includegraphics[width=3in]{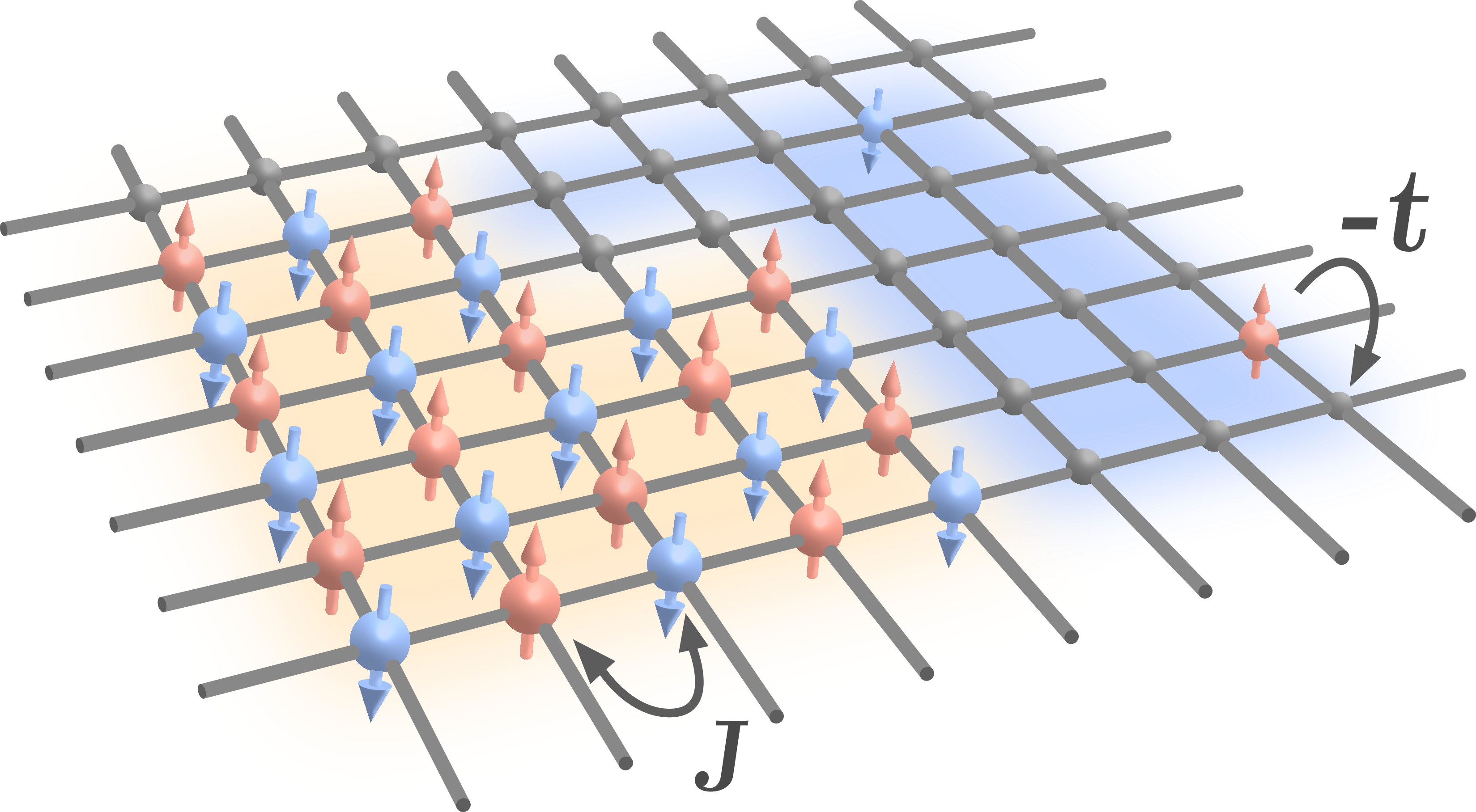}

  \caption[short]{
  \textbf{Initial setup of quench in the $t-J$ model.} 
  We perform an inhomogeneous quench of the chemical potential in the $t-J$ model. The initial state has one half of the system having one electron per site with antiferromagnetic correlations and the other half with a lower electron density. Phase separation in the model leads to this initial state being dynamically frozen for large $J/t$.
  }
  \label{Schematics}
\end{figure}

%----------------------------------------------------------------------%
%  figure 2 - deprecated: ED and MPS time evolution
\begin{figure*}

  \includegraphics[width=\textwidth]{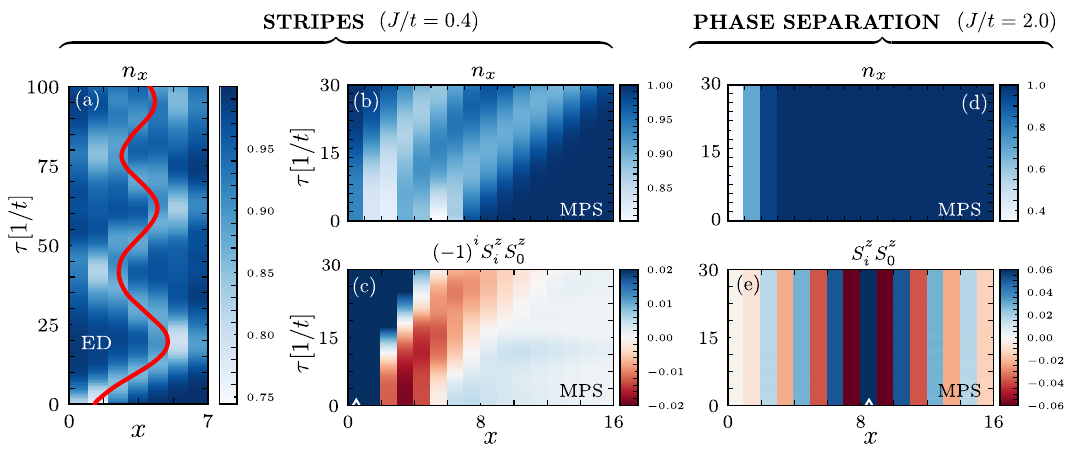}
  \caption{
      \textbf{Dynamics of stripes and phase separation in the $t-J$ model.} (a), (b) and (c) simulate the quench dynamics of stripes in the $t-J$ model for $J/t=0.4$, whilst (d) and (e) correspond to $J/t=2.0$ where the model exhibits phase separation. 
      \textbf{(a)} We use exact diagonalization to compute the rung density, $n_x$ (outlined in main text) of two holes on a $7\times4$ site lattice, subject to a quench of the chemical potential. We observe long coherent oscillations.
      \textbf{(b)} The rung density as in (a) only for an MPS simulation using TDVP on a $16\times4$ site lattice with four holes. Stripes formed on the left half of the system ballistically spread towards the right side of the system. \textbf{(c)} The staggered spin correlations along the longer direction of the cyclinder, for the same simulation as in (b). The domain walls of the antiferromagnet can be seen to move with the maxima in the hole density in (b).
      \textbf{(d)} We simulate the same quench as in (b) only for $J/t=2.0$. The initial phase separated state with holes on one half of the system remains frozen after quenching the chemical potential. 
      \textbf{(e)} spin correlations along the longer direction of the cylinder, which are also seen to remain frozen. The white arrow marks the reference site for the correlations in (c) and (e).
    }

  \label{figure_imshow_timeevo}

\end{figure*}

We use a combination of exact diagonalization using Lanczos techniques on system sizes of up to 28 sites, as well as the time-dependent variational principle (TDVP) method for matrix product states, on system sizes of up to 64 sites in order to establish the dynamical properties of the $t-J$ model subject to an inhomogeneous quench in the chemical potential. Our findings show that a quench for large $J/t$ leads to a completely frozen dynamical regime. We also demonstrate that a quench for $J/t \sim 0.4$, leads to coherent oscillations of stripes for long times. Lastly, we see that the nature of the dynamics for small $J/t$ is largely sensitive to the doping.

%  figure 3: phase diagram and ed time evolution
\begin{figure}

  \includegraphics[width=3.4in]{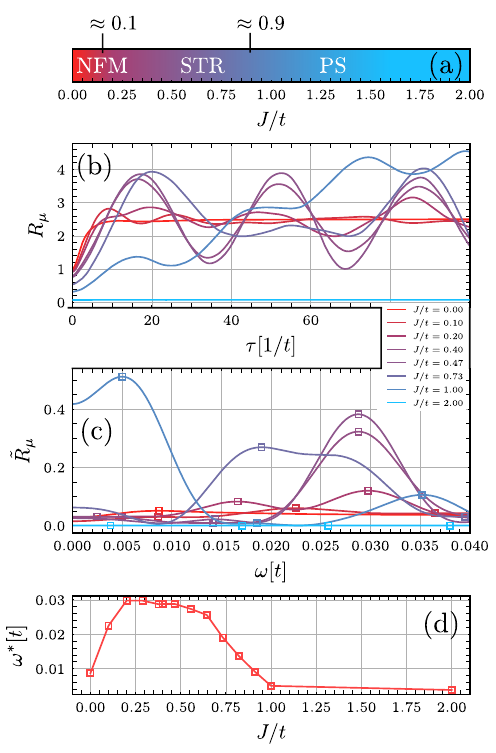}
  \caption{
    \textbf{Phase diagram and center of mass evolution}
    \textbf{(a)} We compute a phase diagram at zero temperature for the 2D $t-J$ model at a doping of 0.125 by using DMRG on a $16 \times 4$ cylindrical geometry (see \cref{appendix A}). At low $J/t$ (red) the ground state develops ferromagnetic correlations, which we label a Nagoaka ferromagnet (NFM). For $J/t > 0.1$ we observe stripes, and for $J/t > 1$ the system phase separates (see \cref{Schematics}).
    \textbf{(b)} ED time evolution of two holes on a $6\times4$ site cluster. We show the center of mass of the two holes, $R_{\mu}$ following a quench (see the thick red line of \cref{figure_imshow_timeevo}(a)). Phase separation leads to a complete freezing of the dynamics. Coherent oscillations of the two holes reflecting from the boundary are seen in the striped phase, and thermalization is seen for small $J/t$ in the NFM phase.
    \textbf{(c)} We compute the Fourier transform, $\tilde{R}_{\mu}$ of the center of mass data from (b) to analyze the frequency of the coherent oscillations and the corresponding hole speed in the striped phase.
    \textbf{(d)} From the Fourier transform of the center of mass data shown in (c), we can compute the dominant frequency, $\omega^*$ for each $J/t$. We note a relatively weak dependence on the speed of the two holes throughout the striped phase. 
  } 
  \label{FigureED}

\end{figure}

%----------------------------------------------------------------------%
%  secion: model
\section{Model}
\label{model}

The Hamiltonian for the $t-J$ model is given by, 

\begin{equation}
  \label{tJ-model}
  H = -t \sum_{\langle i, j \rangle, \sigma} \left(
  \tilde{c}_{i, \sigma}^\dagger \tilde{c}_{j, \sigma} + \text{H.c.} \right)
  + J \sum_{\langle i, j \rangle} \left(
  \bm{S}_i \cdot \bm{S}_j - \frac{ {n}_i {n}_j }{4}\right),
\end{equation}
where the operators $ \tilde{c}_{i, \sigma}^\dagger $ are electron creation operators of spin $\sigma=\uparrow,\downarrow$ at site $i$ that project onto single site occupancies, thereby forbidding doubly occupied sites in the model. $\langle i, j \rangle$ denotes summation over nearest neighbors on a square lattice. $\bm{S}_i=(S^x_i, S^y_i, S^z_i)^T$ are electron spin operators and $n_i$ is the number operator on site $i$.

We study the model in the absence of next-nearest neighbor hopping, where there is no expected pairing in the ground state wave function \cite{doi:10.1073/pnas.2109978118}. We show an approximate phase diagram in \cref{FigureED}(a) at a doping of 1/8 obtained through DMRG calculations; a more detailed discussion of the phase diagram is given in \cref{appendix A}.

%----------------------------------------------------------------------%
%  section: time evolution
\section{Time Evolution}
\label{time_evolution}

We use two computational methods to analyze the real space and time dynamics of holes on a square lattice governed by the $t-J$ model - exact diagonalization (ED) and matrix product states (MPS).
In what follows we study a cylindrical geometry where periodic (open) boundary conditions are employed along the shorter (longer) direction of the lattice. We also introduce a natural notion of left and right into the system, by partitioning the sites with a cut perpendicular to the axis of the cylinder. Times are denoted by $\tau$ in all plots and given in units of the inverse tunneling amplitude, $1/t$.

The initial state for the time evolution is obtained by applying a large chemical potential, $\mu=100t$, to the right half of the system and finding the ground state wave function. We choose an overall hole-doping of $p=1/16$, which effectively translates to a doping of $p=1/8$ on the left half of system. With ED this is performed using Lanczos, whilst for larger systems with MPS we use the Density Matrix Renormalisation Group technique (DMRG). The system is then quenched by switching off this potential and allowing the holes to evolve under the Hamiltonian of \cref{tJ-model}. We consider the rung density, $n_x = \frac{1}{L_y} \sum_y n_{x, y}$ as well as spin correlations, shown in \cref{figure_imshow_timeevo} for several quenches.

We follow Expokit's Krylov-based method in order to time evolve the initial state \cite{10.1145/285861.285868}, in which a Lanczos routine is performed to compute the Hamiltonian with the basis of a Krylov subspace. This smaller matrix can then be exponentiated using a Padé approximation. The time steps are updated based on a local error estimate; we work to within a global error of $10^{-8}$ using a Krylov subspace dimension of $10-30$.
\cref{figure_imshow_timeevo}(a) shows the time evolution of the rung density for a $7\times4$ site ED calculation for $J/t = 0.4$.

To perform time evolution on larger system sizes we use the two-site TDVP algorithm \cite{PhysRevLett.107.070601,Haegemann2016}, and work down to a truncation error of $10^{-7}$. Here the time evolution is limited by a growth of entanglement entropy. The rung density for a simulation on a $16\times4$ site cylinder is presented for $J/t = 0.4$ and $2.0$ in \cref{figure_imshow_timeevo}(b) and (c) respectively. The former case shows a coherent motion of the hole density up to the simulated times of $\tau = 30 / t$. We observe this coherent motion for much longer times in the ED calculation of \cref{figure_imshow_timeevo}(a). The behavior for larger $J/t$ is a completely frozen regime in which the holes remain pinned to one half of the system. Both the ED and MPS simulations could be carried out up to $\tau = 100 / t$ showing no spreading of the hole density.

From the rung density, we then compute the center of mass of the holes during the evolution, 
\begin{equation}
    R_\mu = \frac{L_y}{N_h}\sum_x x \cdot \left(1-n_x \right).
\end{equation}
where $N_h$ is the total number of holes. $R_\mu$ is shown with a red line in \cref{figure_imshow_timeevo}(a) and % This can be seen to trace out the motion of the two holes in both cases. 
for several values of $J/t$  in \cref{FigureED}(b). For large values of $J/t > 1$, this center of mass remains frozen on one half of the system (light blue line). This dynamically frozen regime is also observed in simulations on larger system sizes at the same doping, where long time evolution even with TDVP is possible due to the limited growth of entanglement. For intermediate values of $J/t$ we observe large, persistent oscillations (purple line of \cref{FigureED}(b)), whilst for small values of $J/t$ the center of mass of the hole density reaches equilibrium at the center of the system within a few units of the inverse tunneling amplitude (red line of \cref{FigureED}(b)).

This time evolution indicates a series of interesting dynamical regimes of the $t-J$ model, which we will now proceed to discuss in the context of the ground state phase diagram. We leave a discussion of the convergence of the TDVP results to \cref{appendix B}.

% -------------------------  Groundstate Phase Diagram --------------------------------------------- %
\section{Three Dynamical Regimes}
\label{Phase Diagram}

Performing DMRG on a width four cylinder at doping of $p=1/8$ reveals three phases as we vary the magnetic coupling $J/t$ - a phase-separated state for $J/t > 0.92$; a stripe phase for $0.10 < J/t < 0.92$ and an insulating Nagaoka ferromagnet when $J/t < 0.10$; \cref{FigureED}(a). 

Our ED calculations in \cref{FigureED}(b) indicate that each of these phases is accompanied by a distinctive dynamical regime in performing the quench outlined from \cref{time_evolution}. In this section we combine the longer ED time evolution with that of MPS simulations on larger system sizes and discuss the corresponding dynamics for each of these phases in turn.

\subsection{Stripes}
\label{sub_section_stripes}

We first discuss intermediate values of $J/t \sim 0.4$ relevant to the cuprates, for which we observe a wide range of magnetic couplings where stripes prevail. Stripes were first predicted by Hartree-Fock solutions to the Hubbard model \cite{PhysRevB.39.9749}. They describe a situation in which, the holes situate on the boundary of an antiferromagnetic domain wall.
From our time evolution, we observe the hole density of the stripes to be ballistic in nature up to the times we were able to simulate. For the ED where just two holes are present, the two holes remain locally bound together and oscillate back and forth, reflecting off of the boundary of the cylinder, \cref{figure_imshow_timeevo}(a).

Larger system sizes also indicate a ballistic nature to the motion for the hole density following the quench. An initial state (in the presence of a chemical potential) realizes two stripes on the half of the system with the higher chemical potential. These high hole density regions again remain bound together and spread across the system without mixing for the time scales simulated in \cref{figure_imshow_timeevo}(b). The domain walls of the antiferromagnet can be seen to track the maxima of the hole density as they spread by looking at the staggered spin correlations shown in  \cref{figure_imshow_timeevo}(c), demonstrating coherent motion of the stripes. 

We find the speed at which the holes move and the stripes spread has only a very weak dependency on the magnetic coupling $J/t$. This is seen from the ED time evolution of \cref{FigureED}, in which we first compute the center of mass of the hole density for two holes on a $6\times4$ cluster, $R_{\mu}$, (b).  We Fourier transform this data (with mean subtracted), (c) and take the dominant frequency, $\omega^*$ for different $J/t$, (d). We find this frequency, $\omega^*$, to be relatively constant throughout the striped phase and see the onset of a dominant low-frequency peak as $J/t$ is increased towards the onset of phase separation.

\begin{figure}[h!]
  \includegraphics[width=3.4in]{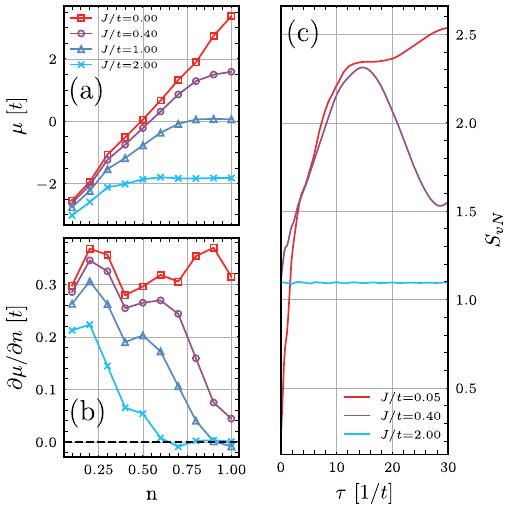}
  \caption{
    \textbf{Maxwell construction and growth of von Neumann entropy.}
    \textbf{(a)} We compute the ground state energy using exact diagonalization as a function of filling, $n$ ($n=1$ corresponds to one electron per site in the $t$-$J$ model). (a) shows the derivative of this energy with respect to the filling giving the zero temperature chemical potential, $\mu$. A flat region of $\mu$ - $n$ curve corresponds to phase separation as seen here for $J/t \ge 1$. 
    \textbf{(b)} Shows the derivative of $\mu$ w.r.t. $n$, which vanishes when phase separation occurs.
    \textbf{(c)} Growth of the of the von Neumann entropy $S_{vN}$ computed during the MPS time evolution of 16 $\times$ 4 cylinder. In the phase separated case ($J/t$ = 2) we note that there is no growth in the von Neumann entropy.
    } 
  \label{Figure_maxwell}
\end{figure}

Lastly we note a decrease of the entanglement during the time evolution, \cref{Figure_maxwell}(c), before increasing again once the hole density reaches the boundary of the cylinder. We show that this result is converged for smaller truncation errors in \cref{MPS time evolution convergence.} of the appendix.

%----------------------------------------------------------------------%
%  figure 4: maxwell svn

%----------------------------------------------------------------------%
\subsection{Phase Separation}

Once $J/t$ is large enough, we cross over from a striped phase into phase separation. Early research into the Hubbard and $t-J$ models raised the question of phase separation and its role in the cuprates. Whilst phase coexistence was not found at parameters relevant to the cuprates, it is present in the $t-J$ model and has a strong consequence for the dynamics. 
We establish phase separation in the $t-J$ model at a doping of $p=1/8$ for $J/t>1$ by means of a Maxwell construction. Using ED calculations on 20 sites, we obtain the chemical potential, $\mu$ for different fillings, $n$ at zero temperature \cref{Figure_maxwell}(a).
Regions of constant chemical potential show the onset of phase coexistence for $J/t > 1$.
Phase separation is also supported with DMRG calculations on larger system sizes with the aforementioned geometry, by computing the charge structure factor and analyzing the charge density, which we leave to \cref{appendix A}.

The phase separation has an extremely strong consequence for the dynamics following the quench we performed - it results in a completely frozen dynamical regime. The rung density, shown in \cref{figure_imshow_timeevo}(d), is static for the duration of the calculated time evolution. This persists even up to times of $100 / t$ where the long time evolution is possible due to a lack of entanglement growth \cref{Figure_maxwell}(c). The holes remain bound to the one half of the system from which they started, failing to mix with the antiferromagnet, for all times we were able to simulate, due to the phase coexisting ground state, tending towards a mutual eigenstate of the system in the presence of a chemical potential imbalance.

\subsection{Nagoaka Ferromagnetism}

The final dynamical regime we would like to discuss lies in the region of small $J/t < 0.1$ of the phase diagram in \cref{FigureED}(a). A theorem due to Nagaoka predicts that a single-hole doped Hubbard model with infinite repulsive interaction (corresponding to $J/t=0$ in the $t-J$ model) admits an itinerant ferromagnetic groundstate \cite{PhysRev.147.392}. We observe the onset of ferromagnetic correlation in the DMRG simulations with domains of length two. 
This leads to the interesting initial state of our quench presenting the interface between a Nagoaka ferromagnet and a Mott antiferromagnet. 
Whilst our ED calculations demonstrate a fast approach to equilibrium in this regime, the absence of ferromagnetic correlation on the small system size leads to a contrast with what we see from the TDVP simulations. 
Here we observe that the ferromagnetic correlations are extremely slow to mix with the antiferromagnetically correlated half of the cylinder and a very slow spreading of the charge degrees of freedom due to the insulating nature of both phases. We therfore leave the MPS time evolution for $J/t = 0.05$ in \cref{NFM.} to the appendix.

%  section: discussion
\section{Discussion and Conclusion}
\label{discussion}

This work has investigated the real-time, real-space quench dynamics of a model paradigmatic to the study of strongly correlated electrons. We have shown using a combination of ED and MPS calculations that a quench in the chemical potential of the $t-J$ can lead to several interesting dynamical regimes. For $J/t \sim 0.4 $, where the ground state shows stripes, we observe long coherent oscillations of the charge density. For calculations on larger system sizes, these oscillations can also be seen in the domain wall of the antiferromagnetic correlations, which can be seen to move coherently with the maxima in the hole density. We also find that the speed at which the stripe spreads has a relatively weak dependence on the magnetic coupling $J$.
Ultimately we expect for long enough times, a thermalization and therefore a decay in the amplitude of said oscillations, but we see that these oscillations remain clear up the long times we are able to compute.

We have shown that large $J/t$, where the ground state is phase separated, leads to a complete freezing in the dynamics. Here there is no growth in entanglement due to the freezing and the initial hole-rich region remains pinned to one half of the system.
Lastly, we find that for small $J/t$, whilst the hole dense region seems to rapidly mix with the antiferromagnetic insulating region on small systems, for large systems, where we start to observe the onset of ferromagnetic correlations, this mixing seems to become inhibited and again we see the two regions coexisting for long times. We leave the exact nature of dynamics in this regime to future studies due to the large sensitivity of the time evolution on exact doping for small $J/t$.

Our results lay a foundation for understanding transport in strongly correlated electron phases as encountered in cuprate superconductors. Recent results~\cite{Sinha2024} point towards a phase separation instability in the two-dimensional Hubbard model in the pseudo-gap regime. As we have demonstrated, phase separation leads to a complete dynamical freezing which in turn could be related to the large resistivity encountered in cuprates in the pseudo gap, strange and bad metal regimes. Moreover, non-equilibrium properties of Fermi Hubbard models can be readily probed using ultracold atoms at intermediate temperatures~\cite{Schneider2012,Langen2015,Scherg2018,Brown2019}. As such our results shine light on the dynamics of the low-temperature ordered phases, which may soon be accessible in these experiments.

\section*{Acknowledgements}
We thank Salvatore Manmana, John Sous, Aritra Sinha, Martin Ulaga, and Roderich Moessner for insightful discussions. A.W. acknowledges support by the DFG through the Emmy Noether program (Grant No. 509755282). The exact diagonalization simulations were performed using the XDiag library \cite{xdiag}. The DMRG and TDVP simulations have been performed using the Julia version of the ITensor library \cite{itensor, itensor-r0.3}.
%----------------------------------------------------------------------%
%  bibliography
%apsrev4-2.bst 2019-01-14 (MD) hand-edited version of apsrev4-1.bst
%Control: key (0)
%Control: author (8) initials jnrlst
%Control: editor formatted (1) identically to author
%Control: production of article title (0) allowed
%Control: page (0) single
%Control: year (1) truncated
%Control: production of eprint (0) enabled
%

%----------------------------------------------------------------------%
%  figure 6: dmrg cyclinder
\begin{figure*}
  \includegraphics[width=.9\textwidth]{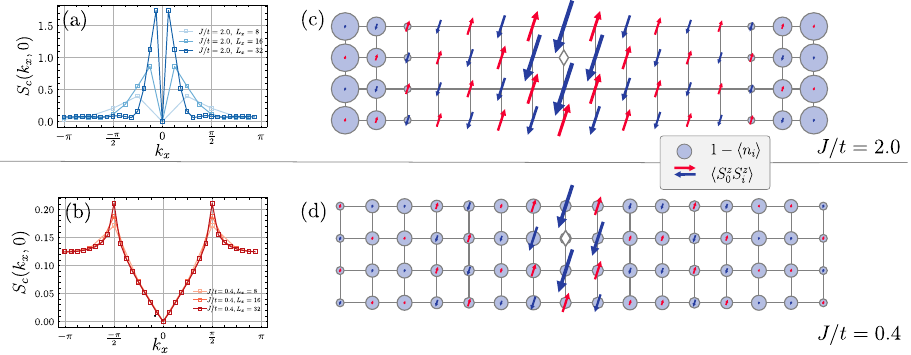}
  \caption{
    \textbf{Phase separation and stripes in the $t-J$ model.}
    \textbf{(a)} and \textbf{(b)} show the charge structure factor at $k_y = 0$ calculated with DMRG for $J/t=2.0$ and $J/t=0.4$ respectively.
    From light to dark shows increasing system size (length of the cylinder). For $J/t=2.0$ (blue) a peak forms at the inner-most non-zero value of $k_x$. This is indicative of phase separation.
    (b) shows the formation of a peak at $k_x = \pi/4$ due to a charge density wave.
    \textbf{(c)} and \textbf{(d)} show the hole density profile and spin correlations for the ground state from DMRG for $J/t=2.0$ and $J/t=0.4$ respectively. The white diamond marks the reference site for the spin correlations. For the phase-separated case, we note that the holes form at the edge of the cylinder, in contrast to the striped phase where a minimum in the hole density is observed at the boundary. }
  \label{Figure: stripes and phase sep}
\end{figure*}

\appendix
%----------------------------------------------------------------------%
%  appendix A
\newpage
\section{Phase Diagram}
\label{appendix A}
\begin{figure}
  \includegraphics[width=3in]{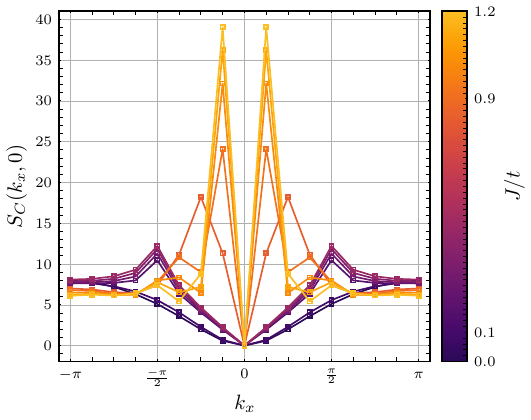}
  \caption{
    \textbf{Charge structure from DMRG.}
    We calculate the charge structure factor at $k_y=0$ from the ground state obtained using DMRG on a $16\times4$ site cylinder. We work down to a truncation error of less than $10^{-5}$ and identify three phases: a commensurate checkerboard insulator, a striped phase and phase separation. We combine this with spin structure factor, hole density distribution and Maxwell construction to estimate the phase boundaries indicated in \cref{FigureED}.}
    \label{Figure_phase_diag_csf}
\end{figure}

In order to identify the phases present in the $t-J$ model at a doping of 0.125 we use DMRG, retaining up to 2000 states and performing 200 sweeps. We measure local electron density $n(x, y)$ as well as density and spin correlations.
We then extract the approximate phase boundary for the transition from a Nagoaka ferromagnet to a striped phase at $J/t \sim 0.1$ by noting a jump in the charge structure factor (see \cref{Figure_phase_diag_csf}). The charge structure factor shows a quadratic dispersion at low momenta for $J/t < 0.1$, which becomes linear for $J/t>0$, suggesting the closing of a charge gap.
For the $J/t > 0.1$ there is also a clear peak in the charge structure factor at a momentum $\pi/2$ due to the wavelength four charge density wave. \Cref{Figure: stripes and phase sep}(b) also shows the charge density and spin correlation on the width four cylinders, which demonstrates the antiferromagnetic correlations with domain walls at the hole maxima - characteristic of a stripe.

The boundary for the transition, from stripes to phase separation is also seen in the charge structure factor, but less clear. Once the peak swaps to the lowest non-zero momentum, the charge density is modulated on a length scale that increases with system size (seen in \cref{Figure: stripes and phase sep}(a)) and the hole density is forced to the edge of the cylinder. This is phase separation between a region of zero hole denisity (center of the cylinder in \cref{Figure: stripes and phase sep}(a)) with strong antiferromagnetic correlations and a high hole density region at the edge of the cylinder. We supplement establishing the boundary of the onset of phase separation by using DMRG at different dopings to perform a Maxwell construction (as in \cref{Figure_maxwell}) and find that the onset of phase separation is for $J/t \sim 1$ on the width four cylinder.

%----------------------------------------------------------------------%
%  figure 7: phase diagram structure factor

%----------------------------------------------------------------------%
%  appendix B
\section{Convergence of results}
\label{appendix B}
\begin{figure}
  \includegraphics[width=3.4in]{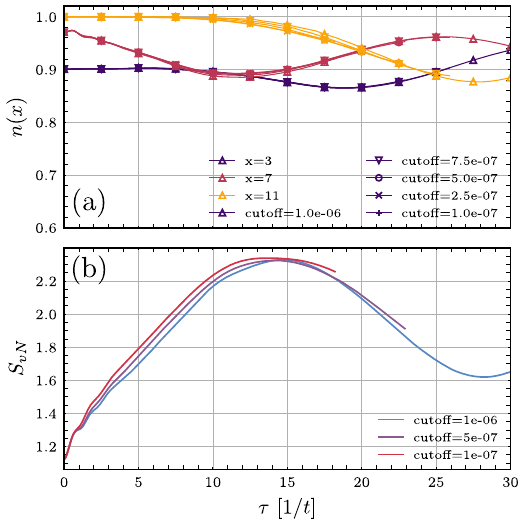}
  \caption{
    \textbf{Convergence of TDVP - $J/t=0.4$}
    \textbf{(a)}
    The onsite electron density as function of time is calculated for cutoffs down to $10^{-7}$ for the TDVP time evolution used in \cref{figure_imshow_timeevo}(c). A bond dimension of 1000 was used to find the initial state and a maximal bond dimension of 2000 was used for the time evolution of the MPS.
    \textbf{(b)} 
    The entanglement entropy plotted for cutoffs from 1e-6 down to 1e-7.
    }
    \label{MPS time evolution convergence.}
\end{figure}

The MPS approximation to the true time evolved wave function becomes a poorer approximation due to the growth of entanglement. For the phase separated case convergence is not an issue, due to lack of growth in von Neumann entropy, whilst for the striped phase there is growth of entanglement. We are however, still able to time evolve with converged results up to times $~30 \frac{1}{t}$. We check for convergence in two ways. We first check that the bond dimension of the initial state (found with DMRG) does not affect the time evolution. We then check for a given bond dimension of the initial state that as we decrease the truncation error in the two site TDVP that the local electron density converges. This is shown in \cref{MPS time evolution convergence.}, where on several sites across the cylinder we check that the density is converged by computing the time evolution down to a truncation error of $10^{-7}$.

\begin{figure}
  \includegraphics[width=3.4in]{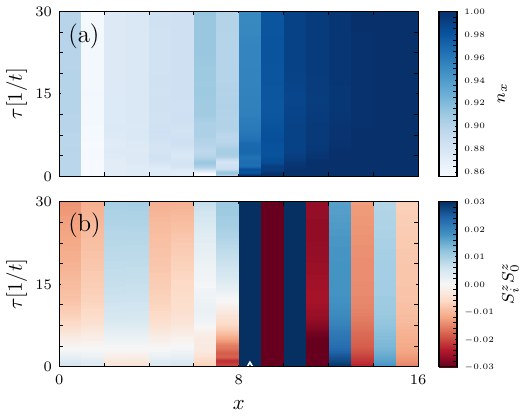}
  \caption{
    \textbf{Dynamics of the Nagaoka ferromagnet}
    Rung density and spin correlations for a quench of the chemical potential for $J/t=0.05$. On the hole doped side of the cylinder, ferromagnetic correlations arise.}
    \label{NFM.}
\end{figure}

\end{document}